# Internal Location Based System for Mobile Devices Using Passive RFID and Wireless Technology

Kapil N. Vhatkar
Computer Technology Department
Veermata Jijabai Technological Institute
Mumbai, India-400 019
kapilnv@gmail.com

G. P. Bhole
Computer Technology Department
Veermata Jijabai Technological Institute
Mumbai, India-400 019
gpbhole@vjti.org.in

*Abstract:* — We have explored our own innovative work about the design & development of internal location-identification system for mobile devices based on integration of RFID and wireless technology. The function of our system is based on strategically located passive RFID tags placed on objects around building which are identified using an RFID reader attached to a mobile device. The mobile device reads the RFID tag and through the wireless network, sends the request to the server. The server resolves the request and sends the desired location-based information back to the mobile device. We had addressed that we can go through the RFID technology for internal location identification (indoor), which provides us better location accuracy because of no contact between the tag and the reader, and the system requires no line of sight. In this paper we had also focused on the issues of RFID technologies i.e. Non-line–of-sight & High inventory speeds.

## I. INTRODUCTION

Rapid advances in wireless technologies and the mobile devices provided an opportunity to develop and deliver new types of location-based application and services to users. Different approaches in location identification can be divided into two main sub categories, internal and external location determination. Cell-ID (Cell Identification), TOA (Time Of Arrival), OTD (Observed Time Difference), GPS(Global Positioning System) are the wireless technologies that are being used for external [5].

However the majority of the current mobile location systems (MLS) lack sufficient accuracy and there are still open challenges with respect to design, usability, functionality and implementation aspects. This research attempts to address some of these issues and present an indoor positioning system architecture based on a combination positioning systems, but GPS is the most used preference for external positioning system, whereas, WLAN (Wireless Local Area Networks), Bluetooth and RFID (Radio Frequency Identification) that are being used for internal positioning system of wireless, RFID and J2ME (Java 2 Micro Edition) technology. The approach has been developed using J2ME.This prototype used a java application with a RFID reader attached to a mobile device in order to locate the user's position in a building with fixed passive RFID tags attached.

In addition the short range of the RFID readers used ensures that the system is able to determine position to high degree of accuracy.

The architecture proposed is flexible and could be deployed in a range of application domains including tourism, inventory tracking and security access. Section 2 reviews some of the related work carried out on location identification and positioning systems. Section 3 provides background information on the RFID technology. Section 4 describes the design, architecture and implementation of the location-based RFID system. Section 5 concludes the paper and draws direction to future work.

## II. RELATED WORK

Existing approaches to location identification can be divided into two main sub categories, external and internal location determination. External positioning systems are usually GPS based (Global positioning systems) or operationally dependant on the augmentation and utilization of existing infrastructure e.g. location of mobile phone masts. GPS uses satellites and works by calculating the time it takes a signal to travel from a satellite to a receiver on a handheld device. Accuracy to within a few meters is achievable using differential GPS. However this approach can be time consuming and unreliable as the GPS receiver needs to be able to communicate with at least four satellites before location can be found. In addition the receiver must maintain a line-of-site transmission with the satellites. As a direct consequence GPS does not work well in built-up areas such as large cities and is not accessible indoors [6].

Different approaches related to internal positioning systems have been used, some of them being briefly discussed in the following.

Active Badge developed at Olivetti research laboratories used infrared technology for indoor location positioning but encountered two major limitations based on line-of-sight requirements and short-range signal transmission [7]. Placelab runs an application on the user's local device to infer current location by using the known longitude and latitude and unique identifier of existing fixed Wi-Fi hotspots, Bluetooth nodes or GSM stations. This observational approach preserves user privacy as the majority of network access points periodically broadcast their presence and monitoring the appropriate frequencies allows these signals to be intercepted and utilized for this purpose. The functionality of Placelab is limited in that it can identify the user's location but then relies on the development of a complementary application to utilize this information effectively. Placelab's accuracy varies widely and is typically within 150 meters using GSM coverage and in the range of 20-50m using Wi-Fi [10].

Due to the limitations of these technologies, RFID has emerged as a more attractive alternative. Radio frequency identification (RFID) is a generic term that is used to describe a system that transmits identity in the form of a unique serial number of an object or person wirelessly using radio waves. This wireless system allows for non-contact reading of RF-enabled tags. RFID can be applied to the development of applications for tourism, libraries, health centres, security access, student tracking, etc. The significant advantage of RFID systems is the non-contact and the non-line-of-sight nature of the technology. RFID systems typically consist of a number of components including RFID tags, RFID readers, antennas and system software. There are several approaches in location-sensing systems that use RFID technology.

SpotON [8] uses RFID technology for three dimensional location sensing based on radio signal strength analysis. Another approach is mTag [3] which is a distributed event-driven architecture for determining location specific mobile web services. The mTag architecture uses fixed RFID readers placed around the building and they are touched with a passive RFID tag attached to a mobile phone or PDA. Our proposal uses a different approach from the mTag architecture which is using fixed RFID tags with RFID readers attached to mobile devices. One of the limitations of this approach is that since it needs specific requirements for execution, it can not be run on any mobile device. LANDMARK [6] approach uses active RFID tags for indoor location sensing. The major advantage of LANDMARK is that it employs the idea of having extra fixed location reference tags to improve the overall accuracy of locating objects. Our proposal uses fixed passive RFID tags which are strategically located around buildings and are identified by an RFID reader attached to a mobile device.and Wal-Mart Stores. RFID describes any system of identification wherein an electronic device that uses radio frequency or magnetic field variations to communicate is attached to an item. Due to the nature of the RFID technology which is very applicable in locating and tracking objects, recently it has been applied to many location identification systems.

## III. RFID TECHNOLOGY

RFID (Radio Frequency Identification) technology is not a new technology, RFID technology has been around for at least 50 years. But the advancements necessary for it to enter the corporate mainstream have only been made recently. The cost has finally sufficiently dropped, and read/transmit range has increased. RFID technology has been used in many organizations and agencies such as the U.S. Department of Defense (DoD), the Food and Drug Administration (FDA), To understand how an RFID tag notifies a reader about its presence and identity, consider the simple scenario depicted in Figure2. In this figure2 the RFID reader transmits radio signals at a preset frequency and interval (usually hundreds of times every second). Any radio frequency tags that are in the range of this reader will pick up its transmission because each has a built-in antenna that is capable of listening to radio signals at a preset frequency (the size and shape of the antenna determine what frequencies it will pick up). The tags use energy from the reader's signal to reflect this signal back. Tags may modulate the signal to send information, such as an ID number, back to the reader. In-short, in a typical RFID System, tags are attached to objects to be technologies that make use of radio waves to automatically identify individual objects.

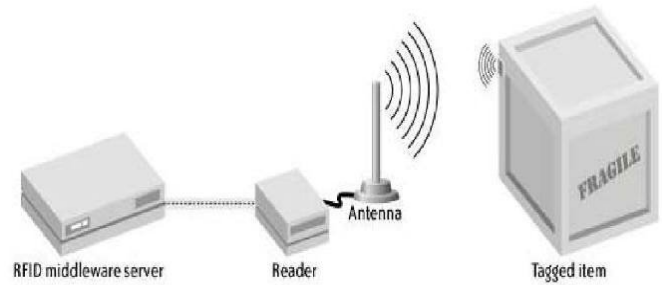

Figure 1:- Components of RFID System [2]

"Radio frequency identification (RFID) is a generic term that is used to describe a system that transmits identity in the form of a unique serial number of an object wirelessly using radio waves."

### A. Components of RFID System

An RFID system is composed of three main elements:
- **RFID Tag** (Transponder): RFID tag or transponder, which usually holds an identification number and is located on the objects to be identified.

- **RFID Reader** (Interrogator): RFID reader or interrogator, which detects tags and reads from and writes to the tags.

- **RFID Antenna**: A coil of wound copper wire, which emits radio frequency signals. It also acts as a receiver.

### B. Working

The RFID tag is located on the objects to be identified and the RFID reader detects tags and reads from and writes to the tags. The reader can then inform another system about the presence of the tagged items. The system with which the reader communicates usually runs software that stands between readers and applications. This software is called as RFID middleware. Each tag has a certain amount of memory to store information about the object, such as its unique tag ID (serial number), or in some cases, more details, e.g. manufacturing date, expiry date etc. When these tags pass through an electromagnetic field generated by the reader, they transmit this information back to the reader, thereby enabling object identification.

When these tags pass through an electromagnetic field generated by the reader, they transmit this information back to the reader, thereby enabling object identification.

*C. RFID Tags*

*1) Types of Tags*

There are two types of tags

**Passive Tags**: Passive tags do not have internal battery and powered by the signal strength emitted by the reader. That means passive tags obtain their operating power from the field generated by the reader.

**Active Tags:** Active tags are larger, more expensive and powered by an internal battery**.**

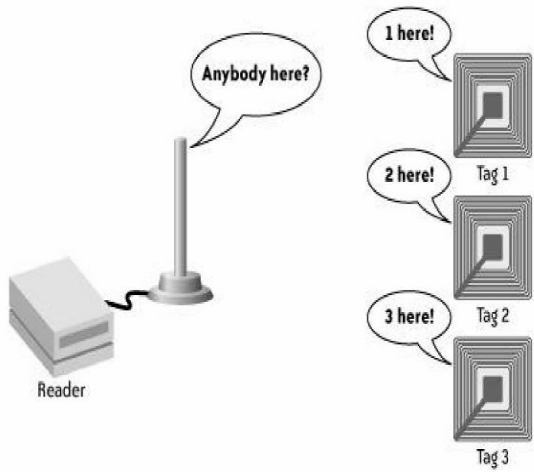

Figure 2:-Communication between RFID tags and reader.

*2) Classes of Tags*

There are four classes of tags:

**CLASS 0** (Read Only Tags) these are the simplest type of tags, where the data, which is usually a simple Tag ID, is stored only once into the tag during manufacture.

**CLASS 1** (Write Once Read Only (WORM)) these can be factory or user programmed. In this case data can then either be written by the tag manufacturer or by the user only once.

**CLASS 2** (Read Write) Data can be read as well as written into the tag's memory. They contain more memory space than what is needed for just a simple ID number.

**CLASS 3** (Read Write with on board sensors) these are active tags which may contain sensors for recording parameters like temperature, pressure etc. and can record the readings in tag memory.

**CLASS 4** (Read Write with integrated transmitters) these tags can communicate with each other without any help from reader.

Basic Tag Capabilities

Many basic operations can be performed with an RFID tag, but only two of them are universal.

**Attaching the tag:** Any RFID tag must be attachable to an item in some way.

**Reading the tag:** Any RFID tag must be able to communicate information over some radio frequency in some way.

Many tags also offer one or more of the following features and capabilities:

Following table1 shows the RFID Tag comparison.

**Table 1:- RFID Tag Comparison**

| Attributes | Active RFID | Passive RFID |
|---|---|---|
| **Tag Power Source** | Internal to tag | Energy Transferred Using |
| **Tag Battery** | Yes | No |
| **Availability of Power** | Continuous | Only In Field Of Reader |
| **Required Signal Strength to tag** | Very Low | Very High |
| **Range** | Up to 100m | Up to 3-5m, Usually Less |
| **Data Storage** | Up to 128Kb | 128bytes |

**Kill/disable:** Some tags allow a reader to command them to cease functioning permanently. After a tag receives the correct "kill code," it will never respond to a reader again.

**Write once:** Many tags are manufactured with their data permanently set at the factory, but a write-once tag may be set to a particular value by an end user one time. After that, the tag cannot be changed except, possibly, to be disabled.

**Write many:** Some tags can be written and rewritten with new data over and over.

**Anti-collision:** When many tags are in close proximity, a reader may have difficulty telling where one tag's response ends and another's begins. Anti-collision tags know how to wait their turn when responding to a reader.

**Security and encryption:** Some tags are able to participate in encrypted communications, and some will respond only to readers that can provide a secret password.

**Standards compliance:** A tag may comply with one or more standards, enabling it to talk to readers that also comply with those standards; or, in the case of standards for physical characteristics, a tag may fit in a particular standard receptacle.

One of its most important attributes is that these tags do not require line of sight to be read. Passive tags are quite smaller and less expensive than active tags. They do not have any batteries and hence obtain their operating power from the reader.

However, passive tags require a higher-powered reader and support shorter read ranges than active tags.

RFID tags are programmable and can store a variety of information including location, destination and product information number. Tags can also be read through a variety of substances such as snow, fog, ice, paint, crusted grime, and other visually and environmentally challenging conditions, where barcodes or other optically read technologies would be useless.

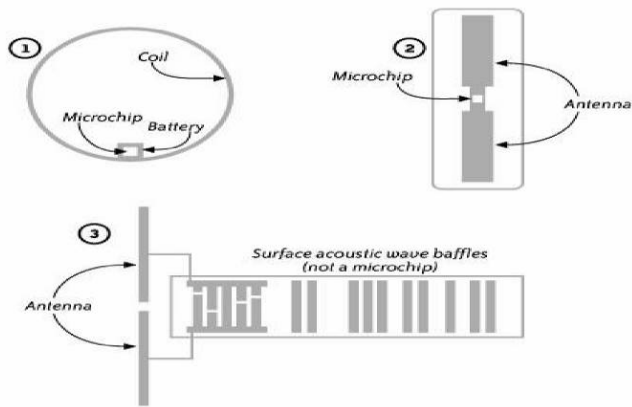

Figure 3.Typical RFID Tags [2]

### D. RFID reader

Since the reader communicates with tags using RF, any RFID reader must have one or more antennas. Because a reader must communicate with some other device or server, the reader must also have a network interface of some sort. Examples of common network interfaces are the serial Universal Asynchronous Receiver/Transmitters (UARTs) or RS 232 or RS 485 communications and the RJ45 jack for 10BaseT or 100BaseT Ethernet cables; some readers even have Bluetooth or wireless Ethernet communications built in. Finally, to implement the communications protocols and control the transmitter, each reader must have either a microcontroller or a microcomputer.

Following figure4 shows the physical components of an RFID reader

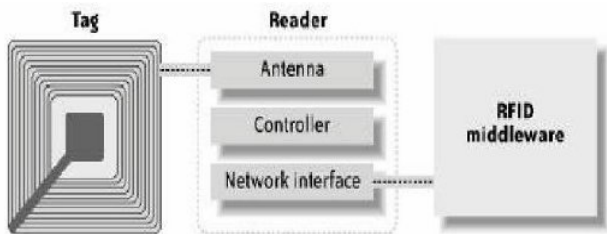

Figure4: Physical Component of Reader [2]

### 1) Types of Reader

Readers, like tags, differ in many ways, and no one reader is a perfect fit for all occasions. Readers come in many shapes and sizes, support different protocols, and often must conform to regulatory requirements, which means that a particular reader may be acceptable for an application in one region of the globe but not in another.

Once identified, a reader may read data from or write to tag memory, depending on the permissions granted by the tag. RFID readers generally fall into two categories-high frequencies (HF) and ultra-high frequency (UHF).Table2 shows a comparison between HF and UHF RFID technology.

Table 2:- Comparison of HF and UHF RFID Technology

|  | HF RFID | UHF RFID |
|---|---|---|
| Frequency | 13.56 MHz | 902-928 MHz N.America<br>860-868 MHz Europe<br>950-956 MHz Japan |
| Read Range | 10-20 cm | 3-6 meters |
| Read Rate | 50 tags/sec | 400 tags/sec |
| Memory Size | 64-256 bit read/write | 64-2048 bits read/write |
| Power Source | Inductive magnetic Field | Capacitive/Electric Field |
| Advantage | Low Cost Standard Frequency | High Speed Longer read range |

The most noteworthy advantage of RFID is that there is no contact between the tag and the reader, and the system requires no line of sight. The tags can be read through a variety of substances and surfaces. They can also be read at extremely high speeds. RFID technology presents many advantages, over other technologies, that signify its suitability for developing internal location-based systems

**Non-line-of-sight:** RFID tags have the advantage that they can be read without line of sight through nonconducting materials. This can save time in processing that would otherwise be spent lining up items.

**High inventory speeds:** Multiple items can be scanned at the same time. As a result, the time taken to count the items drops substantially. RFID tags can be read in less than 100 milliseconds.

**Variety of form factors**: RFID tags come in a wide variety of shapes and sizes. These different form factors allow RFID technologies to be used in a broad range of environments.

**Rewritable tags:** The read/write capability of an active RFID system is also a significant advantage in interactive applications such as work-in-process or maintenance tracking. In the case of reusable container, this can be a big advantage.

If we enter unknown premises it will be very difficult for us to identify a particular shop or a particular area or place in that building. In order to avoid the confusion and difficulties in finding the exact place, we have to get the proper information regarding the different places in that building. To overcome this, passive RFID tags are strategically located around building, for example each classroom/cabin has one passive RFID tag that hold unique identification number. The classroom/cabin's information is stored in the server with the corresponding tag number. When user along with PDA or mobile device (with attached RFID reader) comes nearer to the corresponding tags the area's information will be displayed in the mobile. The user then can send the request to the server, regarding further more information so that some more information from the server will be popped out to the mobile phone as per the user's request.

## IV. ARCHITECTURE & IMPLEMENTATION

The system architecture is illustrated in figure 5. This figure presents an internal location identification system based on the combination of the wireless and RFID technology.

This figure shows the infrastructure and components of the system to be developed. It also depicts the functionality of the application: the active RFID reader with CF (Compact Flash) interface is attached to the mobile device or PDA will detect the RFID tags and will send request through the wireless network to the server. The server resolves the request and sends the answer back to PDA.

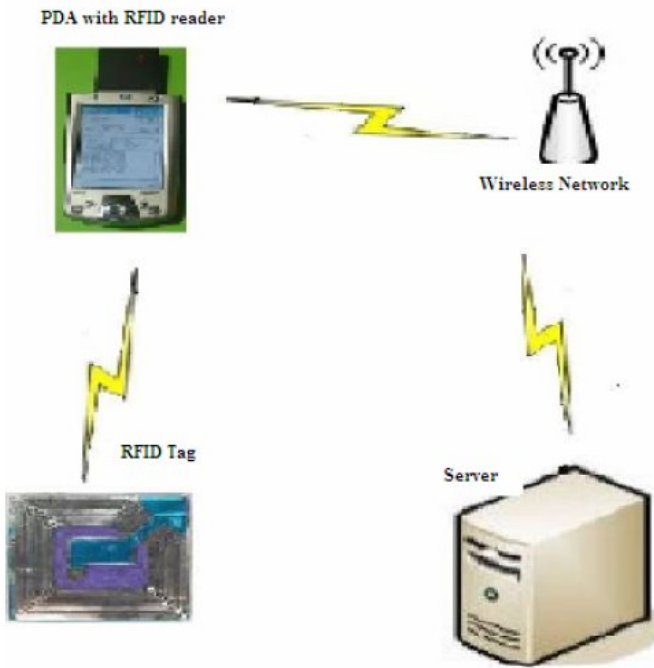

Figure: 5-System Architecture

*A. Implementation Methodology Adapted*

Mobile RFID readers are expensive so that we go for implementation using active stationary RFID Reader (RFID READER-B SL # 702). The communication model is as given below.

Let us see how ILBSFMDURFID System works.

RFID Reader detects the RFID Tag

RFID Middleware takes RFID Tag-Id from RFID Reader and sends it to the Client Application which is running on the PDA/mobile device (J2ME emulator).

Then PDA/Mobile device (J2ME emulator) sends the request to the server through WiFi or other wireless network Server resolves the request and sends answer (i.e. location) back to the PDA/mobile device through WiFi or other wireless network.

The client side contents RFID reader, RFID middleware and client application. The server side contents web server and database.

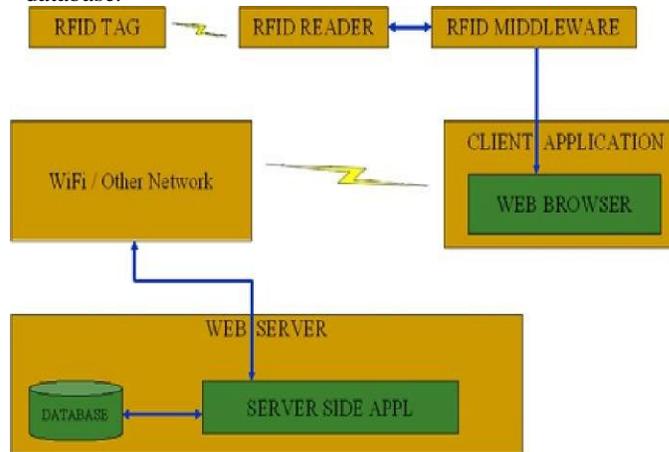

Figure6: The communication model of the system

The client application is developed using Java 2 Micro Edition (J2ME).

Active RFID READER-B SL # 702 that can detect Class 0 tags is connected to the desktop machine through serial TTL interface. The RFID Middleware which is developed using Java takes tag_ID from the reader and gives it to the Client Application running on the The Sun Java Wireless Toolkit which is a software tool that emulates a physical mobile device on a desktop computer. Then client application send request to the server through the wireless network (I used D-Link DIR- 300 wireless G router for the server and D-Link DWL- G510 wireless desktop adapters for the clients). Web server is built using Apache Tomcat and Java Servlets.

Before starting to use this application, the user should connect to the client application. When RFID tag come in the range of the RFID reader then RFID reader reads the RFID tag and client application sends the location ID i.e. tag-ID to the server. The server application then maps the tag_ID with the corresponding entry in the database and retrieves the current location of the user and other related information. Finally, this information is sent back to the user through a wireless network. As a result, the user is presented with information about his/her current location and interest specific info concerning the sight.

*B. Experimental Result*

1. First, RFID Reader detects the RFID Tag.
• The active RFID (RFID READER-B SL # 702) reader is connected to the PC through serial TTL interface.
• This operates in 135 KHz frequency.
• Its read range is 3-4" (10-15 cm).
• It can read Class-0 tag i.e. passive Read Only tags.

2. RFID Middleware takes RFID Tag-Id from RFID Reader and sends it to the Client Application which is running on the PDA/mobile device (J2ME emulator). Following figure shows middleware screen showing tag_ID (110055B53A) detected by the RFID reader.

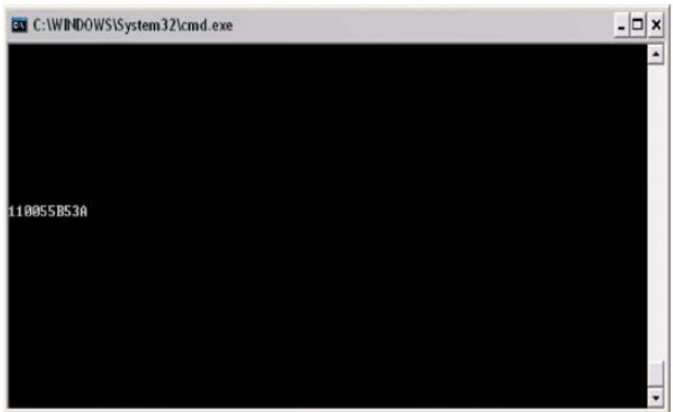

Figure7: RFID Middleware

3. Then PDA/Mobile device sends the request to the server through WiFi or other wireless network, server resolves the request and sends answer back to the user.

4. Figure 8 shows the screen of the mobile emulator: user first need to launch the application.

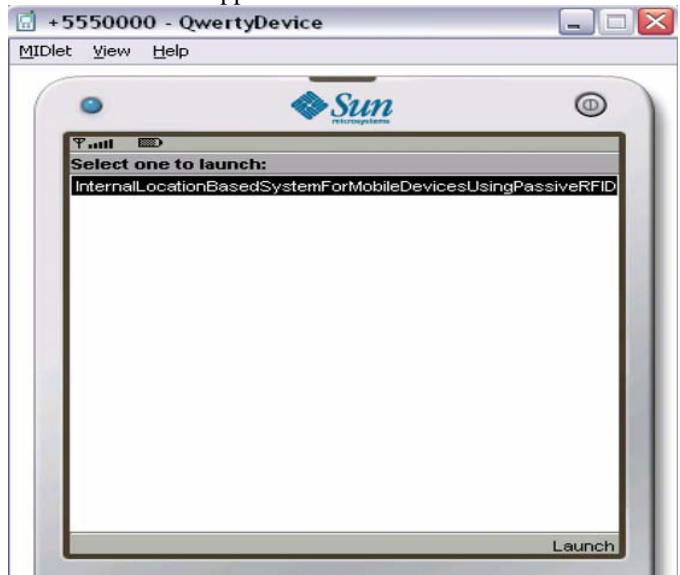

Figure 8. Sun Java WTK and Client Application

5. Following Figure9 shows the login form of the client application, user need to enter username, password and server's IP Address.

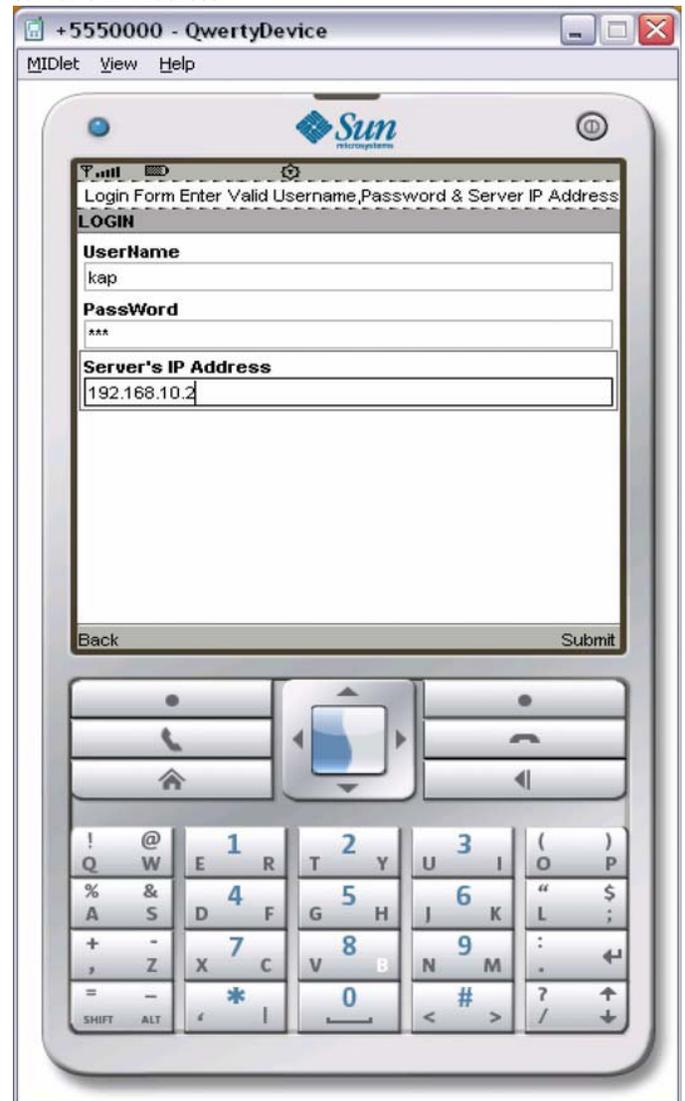

Figure 9. Client Application login page

6. See Figure10:- client application is showing current location of the user. This application checks for update after every 2 seconds that means when RFID Reader detects new RFID Tag (enters into the zone of new RFID tag) then client application updates its form.

7. See figure18 below- Client Application showing sight image.
    In this way our experimental result shows that the system is able to accurately locate user inside the building. This system is based on the integration of J2ME, RFID and wireless technology, using these technologies we were able to provide users with information in different formats including text and images.

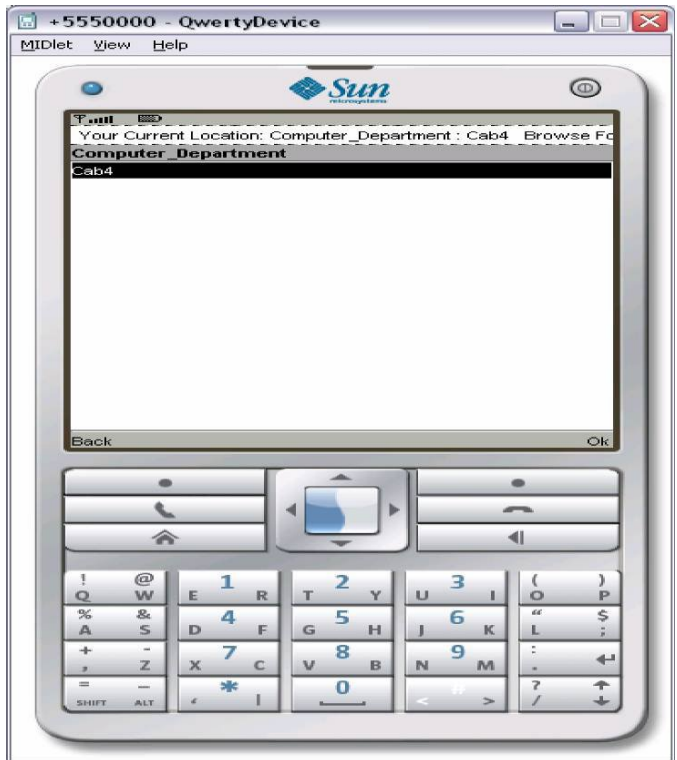

Figure 10. Client Application showing current location of the user

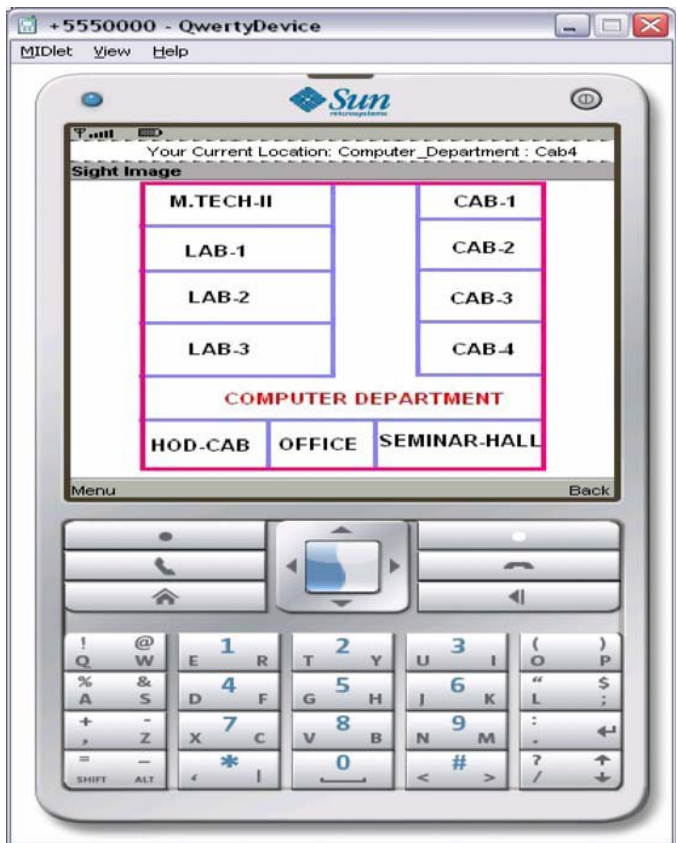

Figure 11. Client Application showing sight image.

## V. CONCLUSION

This paper has presented an architecture and prototype application for delivering internal location-based services for mobile devices using passive RFID technology. The system presented was based on a combination of wireless and RFID technology and was able to accurately locate user and send information based on their location. Using the J2ME for the client application, we were able to provide users with information in different formats including text, image. The most noteworthy advantage of this system is that there is no contact between the tag and the reader, and the system requires no line of sight. The tags can be read through a variety of substances and surfaces. They can also be read at extremely high speeds. Zero power consumption of the passive tags is the key strength of this system. The reduction of price in RFID tags and readers could lead to extensive development of accurate systems and encourage businesses to use it more and more. The problem with RFID systems is that a tag might not be read, in spite of being in the reader's range, due to collisions, this problems need to be resolved to provide efficient solution for tag identification.